\documentclass[]{spie}  %>>> use for US letter paper
\usepackage{amsmath,amsfonts,amssymb}
\usepackage{graphicx}
\usepackage[colorlinks=true, allcolors=blue]{hyperref}
\usepackage{tabularx,multirow}

\title{Status and performance of the THD2 bench in multi-deformable mirror configuration}

\author[a]{Pierre Baudoz}
\author[a]{Rapha\"el Galicher}
\author[a]{Fabien Patru}
\author[a]{Olivier Dupuis}
\author[a]{Simone Thijs}
\affil[a]{LESIA, Observatoire de Paris, PSL Research University, CNRS, Sorbonne Universit\'es, Univ. Paris Diderot, UPMC Univ. Paris 06, Sorbonne Paris Cit\'e, France}

\authorinfo{Further author information: (Send correspondence to P. Baudoz) - E-mail: pierre.baudoz@obspm.fr}

\begin{document} 
\maketitle

\begin{abstract}

The architecture of exoplanetary systems is relatively well known inward to 1 AU thanks to indirect techniques, which have allowed characterization of thousands of exoplanet orbits, masses and sometimes radii.
The next step is the characterization of exoplanet atmospheres at long period, which requires direct imaging capability. While the characterization of a handful of young giant planets is feasible with dedicated instruments like SPHERE/VLT, GPI/Gemini, SCExAO/Subaru and soon with the coronagraphic capabilities aboard JWST, the spectroscopic study of mature giant planets and lower mass planets (Neptune-like, Super Earths) requires the achievement of better coronagraphic performance. 
While space-based coronagraph on WFIRT-AFTA might start this study at low spectroscopic resolution, dedicated projects on large space telescope and on the ELT will be required for a more complete spectroscopic study of these faint planets. 
To prepare these future instruments, we developed a high contrast imaging bench called THD, then THD2 for the upgraded version using multi-DM configuration. The THD2 bench is designed to test and compare coronagraphs as well as focal plane wavefront sensors and wavefront control techniques. It can simulate the beam provided by a space telescope and soon the first stage of adaptive optics behind a ground-based telescope. In this article, we describe in details the THD2 bench and give the results of a recent comparison study of the chromatic behavior for several coronagraph on the THD2.
\end{abstract}


\section{INTRODUCTION}
\label{sec:intro}  % \label{} allows reference to this section

In 2014, GPI \cite{Macintosh08} and SPHERE \cite{Beuzit08} were installed at the Gemini telescope and at the Very Large Telescope (VLT), respectively. They are the first ground-based instruments optimized for exoplanet direct imaging in the near IR (0.95-2.3mic) at separations typically larger than 5 AU and beyond. Therefore, GPI and SPHERE can address the question of planet formation and early stage evolution of planetary system by studying long-orbital-period exoplanets. More importantly, direct observations allow us to study the chemistry of planet atmospheres in the near IR with low spectral resolution. For these reasons, these two instruments can be considered as milestones in the quest for understanding the formation and the atmospheric properties of planets located in the outer part of their system. In addition, imaging cameras in near IR enable disk science, which is sharing the same kind of problematic as exoplanet imaging hence with similar instrumental solutions. Studying directly the connection between disks at various evolutionary stages (protoplanetary to debris phase) is a key in understanding planet formation. At longer wavelengths (2-28mic), the James Webb Space Telescope scheduled for launch in late 2018, will also bring a major contribution to the exoplanet and circumstellar disk sciences based on specialized observing modes. However, because of contrast limitation, which is a challenging aspect of direct imaging, GPI, SPHERE and JWST are sensitive to young gaseous planets ($>$ 1 M$_{Jupiter}$) and more efficient instruments are needed to image older and less massive planets like Neptunes, super-Earths and Earths. These first projects are preparing the scientific grounds and the technologies for the next generation of instruments to realize, eventually, the future of exoplanet science, in particular for what concern the conditions for life.
The closest opportunities to carry out this challenging goal will be offered by the European Extremely Large Telescope (ELT) in the $\approx$2030 era. The instrument PCS (Planetary Camera and Spectrograph) of the ELT will focus on the spectral characterization of mature planets down to a few Earth masses. However, the instrumental concept of such an instrument is still a matter of debate in the community. The final choices will have strong implications on the observational strategy and to the astrophysical results. Therefore, it is crucial to undertake the associated R\&D now. At LESIA/ Observatoire de Paris, we have developed a dedicated high contrast testbed to prepare the next generation of instruments. We emphasize that similar objectives will be carried out in space, in particular with the US mission WFIRST. However, WFIRST being a small telescope we anticipate that only a few atmospheres of mostly giant planets will be studied. Hence, a significant breakthrough is expected from the ELT.

\section{Description of the THD2 bench}

\begin{figure} [ht]
   \begin{center}
   \begin{tabular}{c} %% tabular useful for creating an array of images 
   \includegraphics[height=6cm]{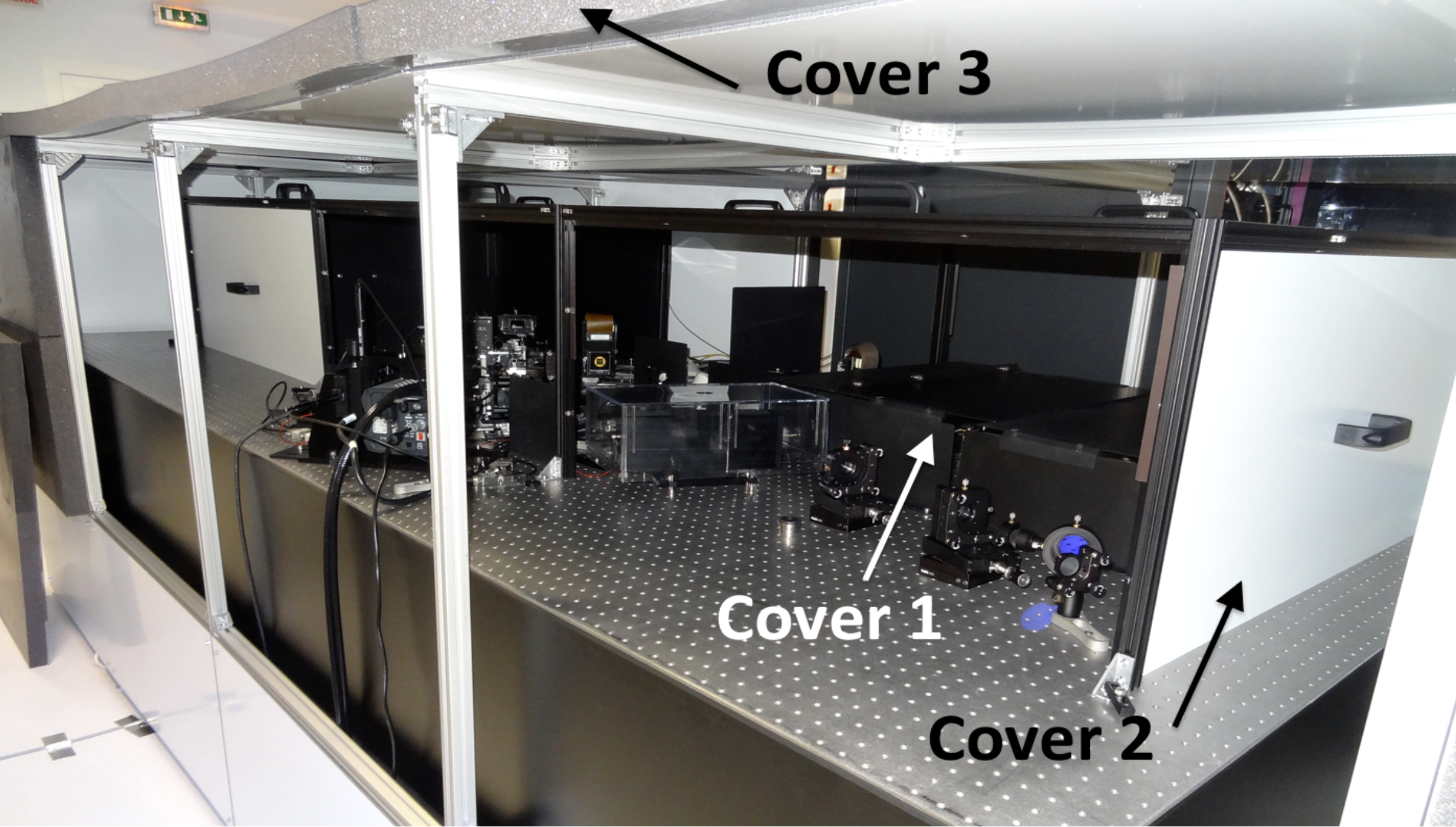}
   \end{tabular}
   \end{center}
   \caption[example] 
%>>>> use \label inside caption to get Fig. number with \ref{}
   { \label{Fig:CoversTHD2} 
Picture showing the covers implemented around the THD2 bench.}
   \end{figure}

The THD2 experiment has been developed at LESIA as an R\&D instrument to study and test high contrast imaging solutions for space-based projects in visible/near-infrared at $10^{8}-10^{9}$ contrast levels and has already achieved some of the best performance worldwide\cite{Mazoyer14, Delorme16_DZPM}. The optical bench is located in an ISO 7 pressurized clean room to limit the dust deposition on the mirrors. Indeed, any particule of dust present in the air or deposited on the mirrors will scatter light and limit the performance level of the coronagraph. The optical elements sit on an optical table that is coupled with a support structures, which use active self-leveling isolators to damp the vibrations of the bench. 
Since temperature gradients can be the source of air movement that creates laboratory turbulence, they have been reduced on the one hand by thermalizing the entire clean room to a precision of 1 degree over several days and on the other hand by creating 3 levels of enclosure around the beam path to stabilize the temperature inside the enclosures to better than 0.1 degree over several hours (Figure  \ref{Fig:CoversTHD2}).

The first enclosure is built around the bench and rest directly on the ground to avoid air flow from the pressurized cleanroom system from exciting vibration modes on the bench. This enclosure is made of aluminum plate on the inside part to allow rapid thermalization of the internal skin of the enclosure. The exterior part is made of foam allowing the attenuation of the residual variation of temperature in the room and the phonic effects. A second enclosure is built on the bench and is made of a thin foam sandwiched between aluminum plates. This second enclosure protect the full beam path except the science detector and the injection module. 

Assuming the beam is located in a box with small height compared to the horizontal dimension, the convection inside the enclosure can be described by a B\'enard-Rayleigh convection. In this case, the convection cells produced within an enclosure have an energy proportional to $h^3$, with $h$ the height of the box. To minimize the height of the box, we chose optical and mechanical elements allowing the beam to propagate close to the table (85 mm) and added a third structure made of horizontal and vertical metal structures arranged on the bench as close as possible to the beam path to cut the effects of turbulence. To limit human presence in the clean room, all essential movable elements are remotely controllable via motors.

   \begin{figure} [ht]
   \begin{center}
   \begin{tabular}{c} %% tabular useful for creating an array of images 
   \includegraphics[height=10cm]{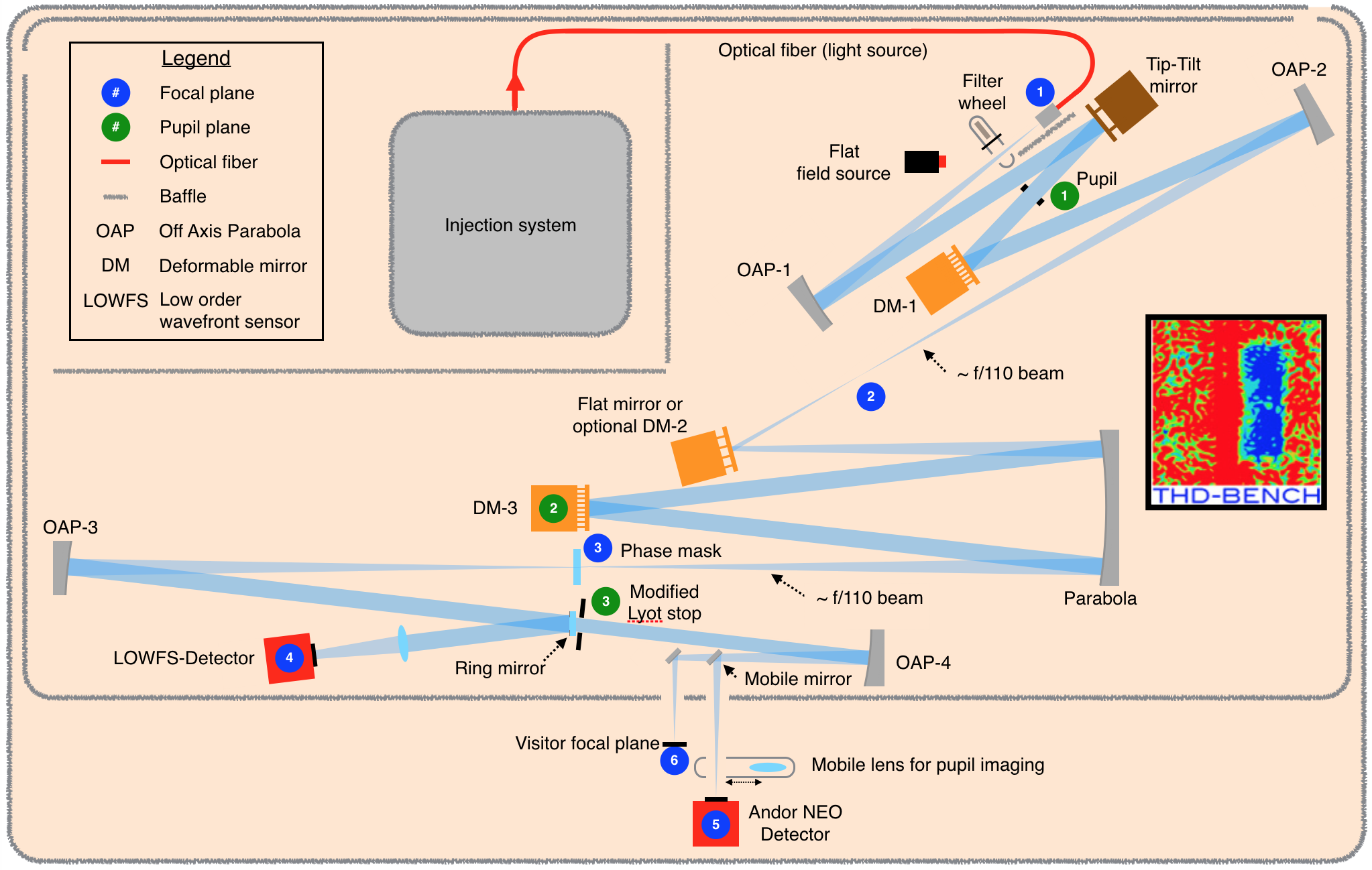}
   \end{tabular}
   \end{center}
   \caption[example] 
%>>>> use \label inside caption to get Fig. number with \ref{}
   { \label{Fig:DesignTHD2} 
Physical implementation of the elements on the THD bench.}
   \end{figure} 
   
The THD2 design is shown in Figure \ref{Fig:DesignTHD2}. The dimensions and orientation of the various elements are generally respected. The beam path (in blue) was intentionally enlarged. The off-axis parabolas and the parabola are cut from full parabolas made of Zerodur. The mirrors have a roughness of the order of a few nm and mechanical surface quality estimated at $\lambda$/20 over an area of 400 mm$^2$. A light injection module (Section \ref{sec:Injection_THD2}) located between enclosure 1 and enclosure 2 feeds the bench via a monomode optical fiber which acts as an angularly unresolved point light source. It is fixed and serves as a reference for the alignment of the bench. At its output the light propagates in a divergent beam. It first passes through a filter wheel and then is collimated by an off-axis parabola (OAP-1) which reflects the beam towards a tip-tilt mirror (Section \ref{sec:Active_THD2}) located 90 mm from the first pupil plane of the bench. This entrance pupil is mounted on a stage that allows precise positioning and focusing of the pupil. Four different circular pupils with a maximum diameter of 9mm are available on this stage. Other pupil shapes can be implemented easily if required. The beam can be apodized at this level by putting an apodizer very close to the pupil. After this pupil plane, the beam is intercepted by a first 34x34 actuators Deformable Mirror (DM-1) which is located at a distance of 250 mm from the pupil plane, then a second off-axis parabola (OAP-2) which focuses the beam in an unused focal plane. The beam is then reflected on a flat mirror that can be replaced by an optional 12x12 DM (DM-2) that is located in a diverging beam. The equivalent optical distance of this optional DM to the pupil plane is 2.5m and is designed to correct low-order amplitude aberrations. The beam is sent to a parabola that creates a pupil plane where a 32x32 DM is located (DM-3). DM-3 returns the beam to the parabola which converges it in the second focal plane of the bench. Different coronographic phase masks can be inserted manually in this focal plane using high-precision repositioning kinematic mount. The light is then collimated by the third off-axis parabola (OAP-3) to create a third pupil plane where the Lyot module is located. The last off-axis parabola (OAP-4) focuses the beam to a planar mirror which folds it outside of the enclosure 2 where two focal planes are available. One focal plane harbors a NEO s-CMOS detector from ANDOR while the other is a visitor focal plane where different instruments can be installed. Pupil plane imaging is also offered by inserting a lens in the beam using a motorized stage. The detector is a 2560 x 2160 pixels detector but we usually use a small region of interest of 400x400 that allows a frame rate of 100 Hz. The pixel size is 6.5 $\mu$m with a readout noise of about 2e- and a full well capacity of about 15 000 e-.

\subsection{Active mirrors}
\label{sec:Active_THD2}
The bench includes 1 tip-tilt mirror and 3 deformable mirrors.
The tip-tilt mirror is one of several prototypes that where developed in the context of SPHERE. It has good linearity ($<2\%$) and its resolution is better than 0.16 $\mu$rad rms with a total stroke of 2 mrad. Its maximum frequency is about 100Hz. DM-1 is a Boston Micromachines 952 actuators deformable mirror (34 across a circular aperture of 9.9 mm) with a maximum stroke of 600 nm. DM-3 is a 32x32 deformable mirror with an actuator pitch of 300 $\mu$m and a maximum stroke of 1500 nm. This last mirror sets the maximum pupil diameter for the bench to 9.3mm. The multi-DM configuration of the bench was developed to allow correction of both phase and amplitude aberrations using Fresnel propagation to introduce amplitude effects with deformable mirrors located outside of the pupil plane \cite{Shaklan06}. This multi-DM configuration will also be used to study the implementation of chromatic corrections on the bench. The distance between the pupil and DM-1 has been set to 25 cm to allow correction of the highest frequencies of the amplitude aberrations. To avoid large DM stroke and their related non-linearity effects, low-order amplitude aberrations requires a DM  placed at a larger distance from the pupil. To limit the size of the bench, we plan to implement in 2018 a DM in a diverging beam with a effective distance to the pupil of 2.5m. Since the beam is smaller at the DM-2 location than for DM-1 and DM-3, we need a DM with a smaller active surface. DM-2 is a 12x12 deformable mirror with an actuator pitch of 300 $\mu$m and a maximum stroke of 1500 nm. Assuming a pupil of 8mm on DM-1 and DM-3, the beam on DM-2 will have a diameter of 2.4 mm which corresponds to 8 actuators across the beam. Since, it is mostly optimized for the correction of low-order amplitude aberrations, the low number of actuators across the pupil should not be an issue.

\subsection{Injection module}
\label{sec:Injection_THD2}
The injection module consists of a series of three lasers centered at 637 nm, 705 nm and 785 nm that are combined using two dichroic fiber combiners and a supercontinum source coupled with bandpass filters. The transition from one to the other is done by rotating a flat mirror to inject the useful light into a single mode fiber. A small part of the light (1$\%$) injected into this fiber is used to measure simultaneously the total injected photometry and the spectrum of the source. The photometry is measured with a photodiode sensitive between 350 nm and 1100 nm, with a resolution of 10 pW, and a dynamical range of 5 $10^8$. The spectrometer is an Ocean Optics USB4000 with a resolution of 1nm to 2 nm with a sensitive range between 400 and 900 nm. Measurements from the photodiode and from the spectrometer are recorded simultaneously with the NEO camera images during operations.

   \begin{figure} [ht]
   \begin{center}
   \begin{tabular}{c} 
   \includegraphics[height=5cm]{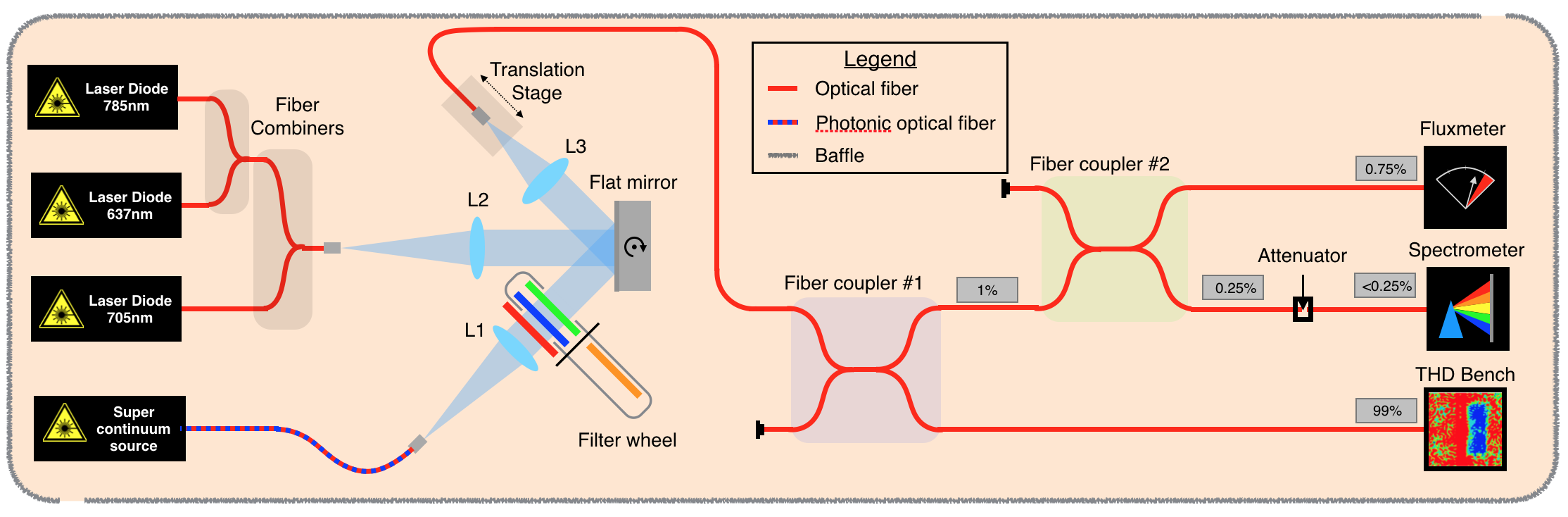}
	\end{tabular}
	\end{center}
   \caption[example] 
   { \label{fig:Injection_THD2}
Design of the THD2 injection module}
   \end{figure}    
 
\subsection{Lyot stop module}
\begin{figure} [ht]
   \begin{center}
   \begin{tabular}{c} %% tabular useful for creating an array of images 
   \includegraphics[height=6cm]{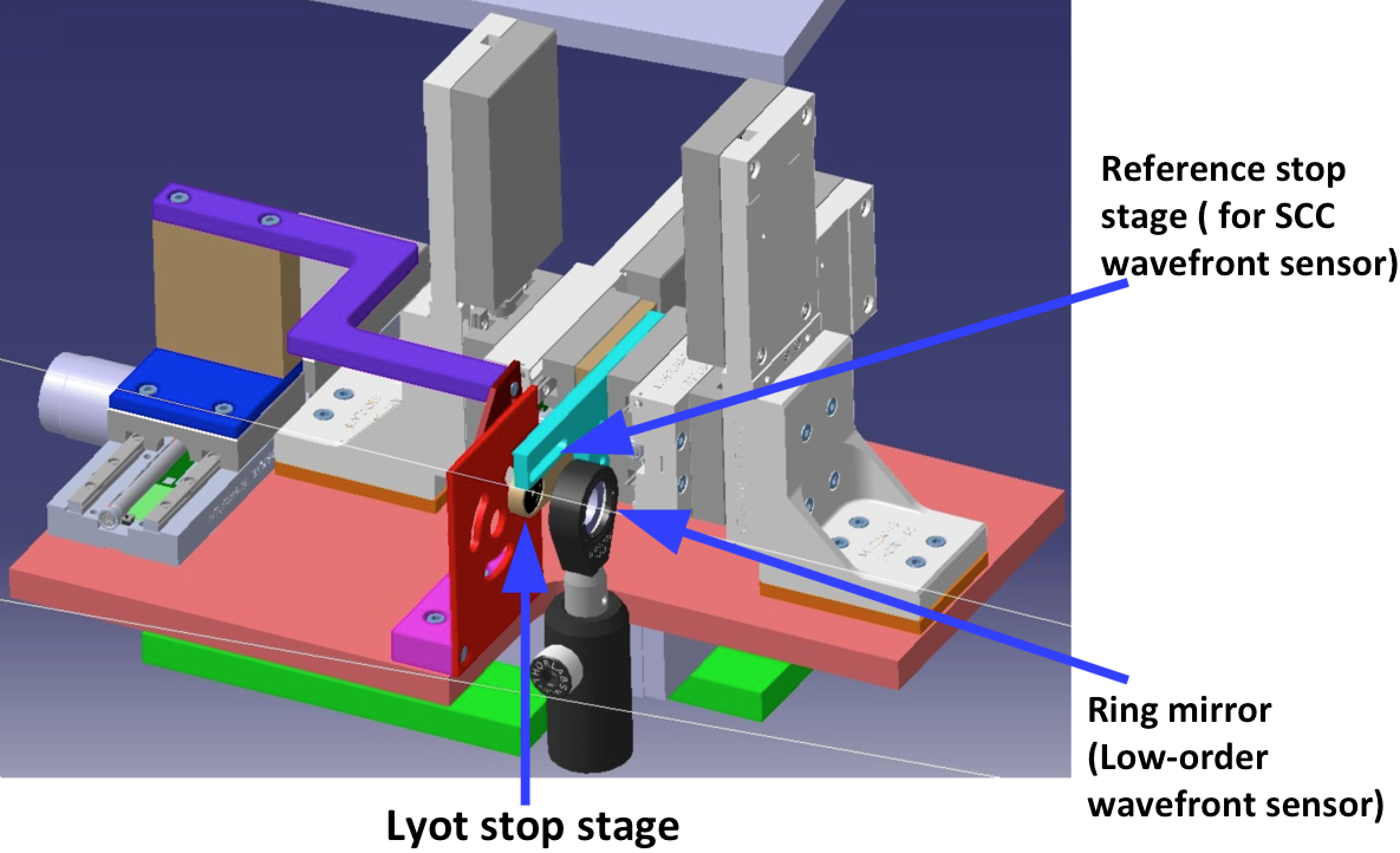}
   \end{tabular}
   \end{center}
   \caption[example] 
%>>>> use \label inside caption to get Fig. number with \ref{}
   { \label{Fig:Lyot_stage_THD2} 
Physical implementation of the elements in the Lyot stop module.}
   \end{figure}

The Lyot stop module consists of a Lyot stop stage, a stage to create a reference hole to allow the use of the Focal plane wavefront sensor called Self-Coherent Camera and a ring mirror used for the low-order wavefront sensor (Figure \ref{Fig:Lyot_stage_THD2}). The Lyot stop stage allows choosing between 4 different Lyot sizes (6.5, 7.9, 8.0 and 8.1 mm by default) and other diameters can be implemented. The reference stage allows us to change the size of the reference hole mandatory for the use of the Self-Coherent Camera (Section \ref{sec:WFS}). Hole size varies from 2mm to 0.3mm. The ring mirror is made of a 20 mm diameter mirror with a central hole of 12 mm diameter. It reflects the light diffracted outside of the Lyot stop by the focal plane mask into the low-order wavefront sensor arm (Section \ref{sec:LOWFS}).

 \subsection{Low-order wavefront control loop}
\label{sec:LOWFS}
The low-order wavefront sensor we are using is based on the concept of the Lyot-based low-order wavefront sensor (LLOWFS) proposed by Singh et al. 2014\cite{Singh14}. This LLOWFS is a coronagraphic wavefront sensor which is designed to sense the pointing errors and other low-order wavefront aberrations. The implementation on the THD2 uses the light diffracted outside of the geometrical pupil downstream of the focal plane mask. This light is extracted from the beam by a ring mirror placed right before the Lyot stop. The light is then focalized by a lens on a CCD camera, which is a 640 pixels x 480 pixels AVT Pike detector with a 7.4 $\mu$m pixel size and a readout noise of 18e-. Frame rate for the full detector is 100 Hz and can be increased to 1000 Hz when reading only a region of interest of 30 x30 pixels, which is enough for the low-order loop. Up to now, we have only measured the tip-tilt with our our low-order loop. We calibrate the sensor by recording the 30 pixels x 30 pixels image introducing tip, tilt and subpixel shifts of the image to compensate for camera movement. We apply a generalized inversion of the matrix created with these recorded data to create a control matrix. We apply this last matrix to correct the tip-tilt and the drift of the LLOWFS camera position. This control loop is especially useful to avoid slow drift of the beam with respect to the coronagraph axis when recording interaction matrix of the focal plane wavefront sensor. The effective bandwidth of the loop is limited by the tip-tilt mirror bandwidth to about 100 Hz.

\section{Focal plane wavefront sensing and correction}
\label{sec:WFS}
To reach an efficient Dark Hole correction \cite{Borde06} with the active mirrors, a focal plane sensor is required to estimate the full electric field (phase and amplitude) at the detector focal plane. For this purpose, we are routinely using the Self-Coherent Camera (SCC\cite{Baudoz06}). The SCC use the light diffracted in the Lyot plane by a focal plane mask coronagraph. A small pupil, called reference pupil, is added in the Lyot stop plane to select part of the stellar light that is diffracted by the focal coronagraphic mask. The two beams are recombined in the focal plane, forming Fizeau fringes, which spatially modulate the speckles created by the aberrations upstream of the coronagraph \cite{Galicher08}.. These encoded speckles in the SCC images allow us to retrieve the complex amplitude in the final coronagraphic image, construct a control matrix for our SCC loop and correct for the aberrations in monochromatic light \cite{Mazoyer14} as well as in polychromatic light \cite{Delorme16_MRSCC}. More information on the process we apply to record the interaction matrix and construct the control matrix for the SCC is given in Baudoz et al. 2012 \cite{Baudoz12}. Using a model of the coronagraph, we can also reconstruct the aberrations in the entrance pupil as shown in Mazoyer et al. 2013\cite{Mazoyer13}.
We are also starting to test other wavefront sensing solutions based on an extension of phase diversity to coronagraphy \cite{Sauvage12,Herscovici17}.

\section{Comparison of several coronagraph spectral performance}

THD2 has proved to reach the same results than THD in monochromatic performance with contrast of $\approx 10^8$ reached in monochromatic on Half Dark Hole between 5$\lambda/D$ and 12$\lambda/D$ \cite{Mazoyer14}. In polychromatic, we have shown that the contrast was degraded by a factor up to 6 for large spectral bandwidth (37\%) \cite{Delorme16_DZPM}.
To prepare the future instrument of characterization of the exoplanets it is crucial to develop coronagraphs, focal plane wavefront sensors and correction strategies compatible with large spectral bandwidths. To define the best solution for large spectral bandwidth, we have started a comparison study of the spectral performance between different types of coronagraphs on the THD2. We compared between January 2017 and March 2017 four types of coronagraphs:
\begin{itemize}
\item FQPM: Four Quadrant Phase Mask \cite{Rouan00}
\item DZPM: DualZonePhaseMask \cite{Soummer03}
\item SLPM: Six Layers Phase Mask \cite{Hou14} which is an achromatization of the FQPM
\item Vector Vortex based on liquid crystal polymers \cite{Mawet09}
\end{itemize}

   \begin{figure} [ht]
   \begin{center}
   \begin{tabular}{c} %% tabular useful for creating an array of images 
   \includegraphics[height=6cm]{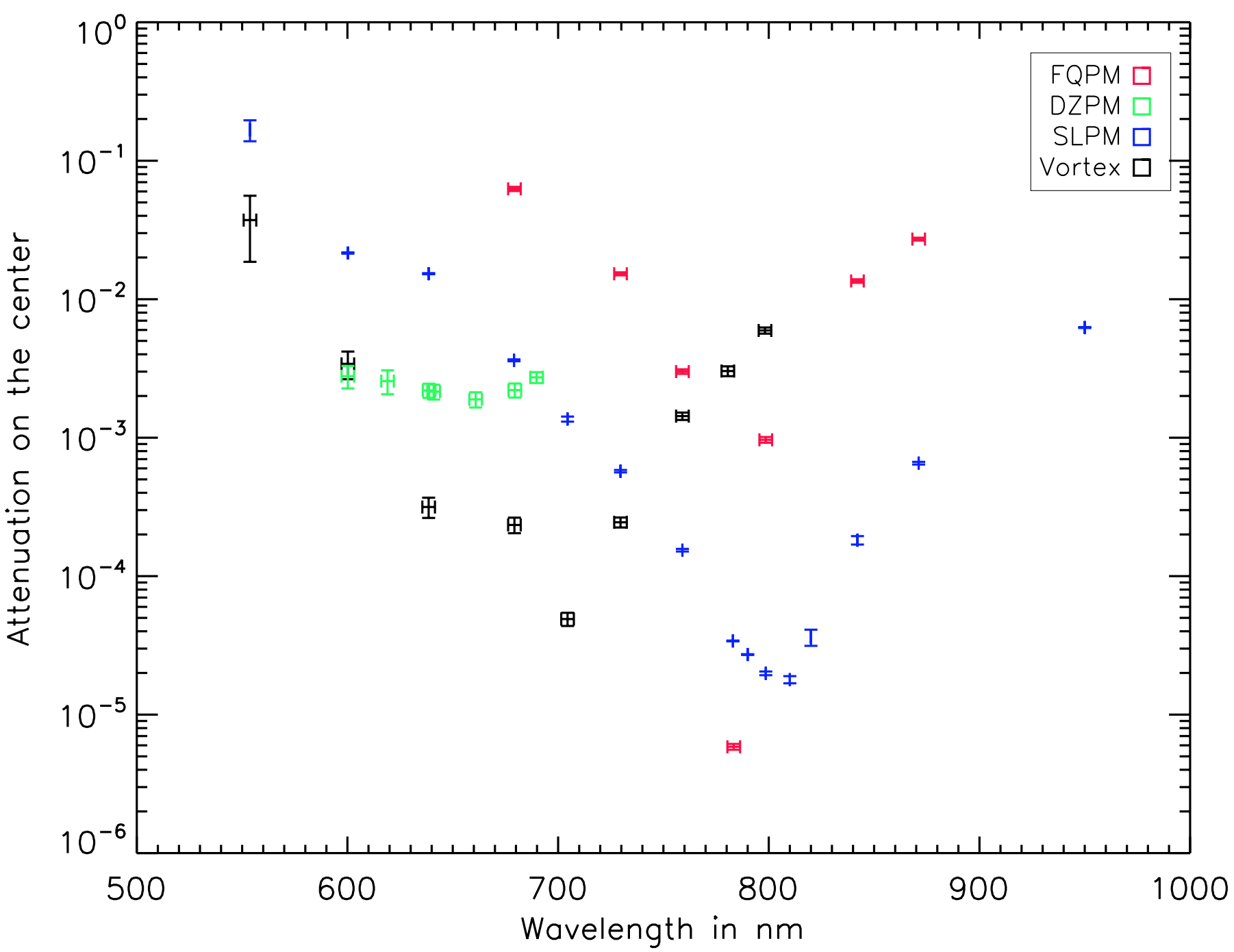}
   \includegraphics[height=6cm]{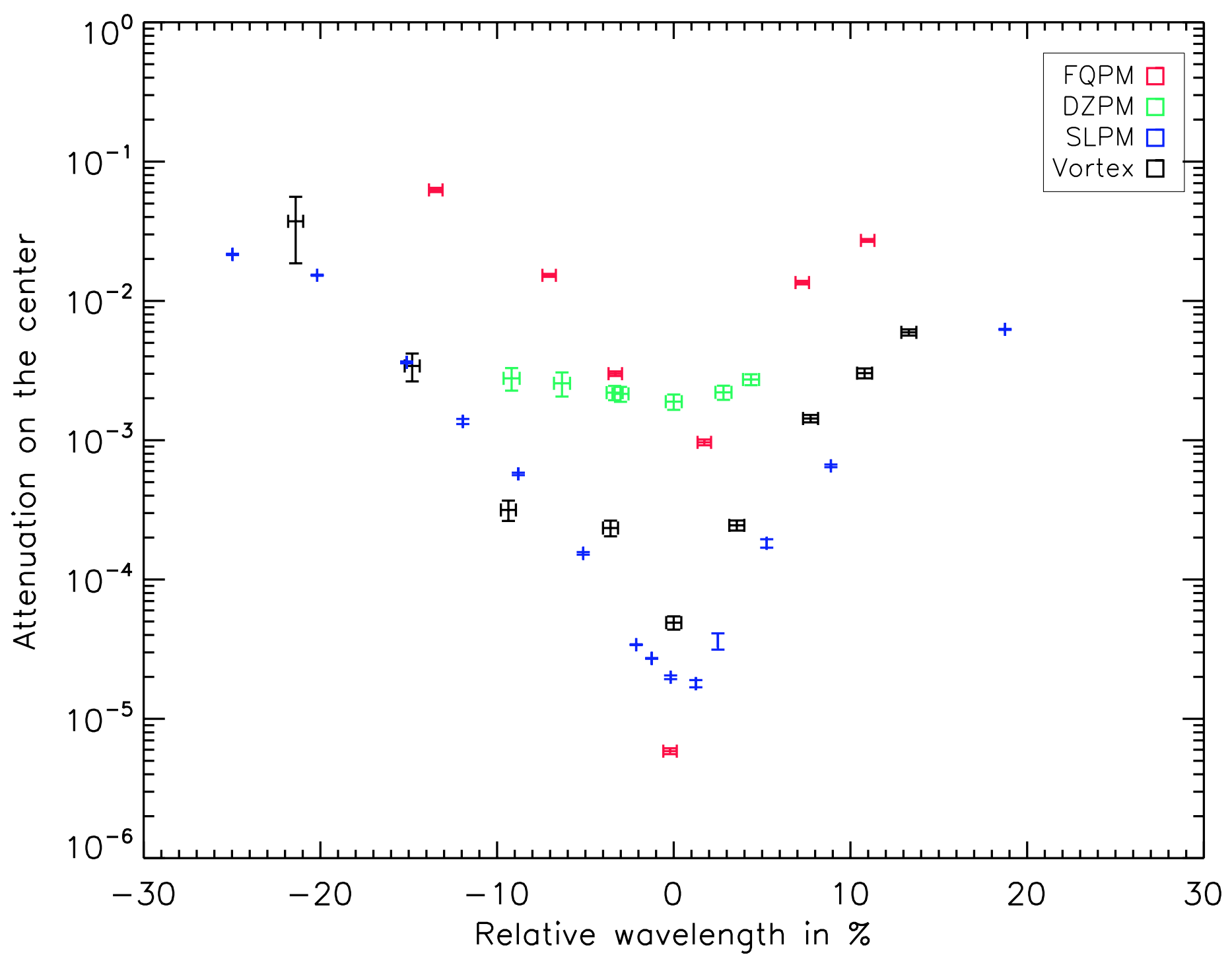}
   \end{tabular}
   \end{center}
   \caption[example] 
%>>>> use \label inside caption to get Fig. number with \ref{}
   { \label{Fig:coro} 
Left: Attenuation (ratio of the maximum of the images on-axis and off-axis of the coronagraph) as a function of wavelength for different wavelengths and coronagraphs. Right: Attenuation as a function of the wavelength relative to the optimized wavelength of the coronagraph.}
   \end{figure} 

To compare their intrinsic chromatic limitation, we are measuring the attenuation between the image of the source when centered on the axis of the coronagraph and when located at a distance of about 7 $\lambda/D$. We are measuring the ratio of the maximum of both images for a series of wavelength and different types of coronagraphs (Figure \ref{Fig:coro}, Left). All the spectral measurements are recorded with spectral bandwidth of about 10 nm using the supercontinum source except for the 3 wavelengths 637 nm, 705 nm and 785 nm where we use the lasers to record the data. First, we notice that the coronagraphs are not optimized for the same central wavelength with the DZPM mostly built to work around 637 nm, the vortex centered around 700nm and the FQPM and the SLPM optimized around 780-800 nm. These optimizations are partly historical since the THD was first working around 637 nm and we slowly increased the spectral bandwith up to 800 nm. The measurement have been done after a full Dark Hole correction at the optimized wavelength for each coronagraph. Then, we apply the same shapes on the DMs for all the wavelengths to perform the measurements (as described in Delorme et al. 2016 \cite{Delorme16_DZPM}). We also checked in a few cases the effect of correction at a different wavelength to verify that the attenuation level measured was really linked to the coronagraph limitations and not set by chromatic aberrations from the bench. 

To better compare the different coronagraphs, we rescaled the measurements to the same wavelength scale relative to the central wavelength of each coronagraph (Figure \ref{Fig:coro}, Right). This is better suited to compare the chromaticity of each coronagraph. We can see that the FQPM is reaching the best monochromatic results but is very rapidly limited when the spectral bandwidth is larger than a few percent. The DZPM we have been using is very achromatic but is relatively limited in attenuation. SLPM and vortex reach about the same level with an attenuation better than $10^{-3}$ for spectral bandwith of about 30\%. It starts to be sufficient to perform study of the chromatic behavior of focal plane wavefront sensors and correction strategies. Note that the results with the vector vortex we are using was recorded without polarizing elements. Using the vector vortex between circular polarizers gave better results but the curve is not given here.

\section{Collaboration}

Several of these results and more to come are based on collaborations between our institute (LESIA Observatoire de Paris) and national or international partners. We recall in Table \ref{tab:collab} the different collaboration we gathered around the THD2 bench in the last few years.

\begin{table}[ht]
\small       
\begin{tabularx}{17cm}{|X|X|X|l|} 
\hline
   & Institutes & Study description & dates               \\
\hline
\multirow{6}{*}{Coronagraphic components} & LESIA-GEPI Obs. Paris, France & Four Quadrant Phase Mask, Multi-Four Quadrant Phase Mask & 2010-2016\\
						\cline{2-4}
						& Laboratoire d'Astrophysique de Marseille (LAM), Marseille, France & Dual Phase Mask coronagraph & 2012-2016\\
						\cline{2-4}
						& Hokkaido University, Japan & Achromatic Eight Octants Phase Mask coronagraph based on photonic crystal & 2015-   \\
						\cline{2-4}
						& National Astrophysical Observatory of Japan (NAOJ) & Vortex coronagraph based on photonic crystal & 2015-   \\
						\cline{2-4}
						& LESIA Obs. Paris, France & Achromatic vortex coronagraph based on liquid crystal polymers  & 2017-   \\
						\cline{2-4}
						& Laboratoire Ondes et Mati\`ere d'Aquitaine (LOMA), Bordeaux France & vortex coronagraph based on electro-optically addressed liquid crystal thin films & 2017-   \\
\hline
Focal plane wavefront sensor estimation and correction 	 &	 LESIA Obs. Paris & Monochromatic and polychromatic Self-Coherent Camera (SCC) used to study multi-DM correction of phase and amplitude simultaneously and achromatic correction & 2010- \\
						\cline{2-4}
						& The French Aerospace Lab (ONERA) & Coronagraphic phase diversity (COFFEE) & 2015-   \\
						\cline{2-4}
						& Netherlands Institute for Space Research (SRON)  & Optimization of chromatic correction with Multi-DM correction & 2015-   \\
						\cline{2-4}
						& LESIA Obs. Paris, France & Focal wavefront sensor and correction in the presence of residual atmospheric turbulence & 2018-   \\
\hline

\end{tabularx}
\vspace{12pt}
\normalsize
\caption{Lists of studies underway on the THD2 and related collaborations.}
\label{tab:collab}

\end{table}

\section{Future work and conclusion}
The THD2 experiment has been developed at LESIA as an R\&D instrument to study and test high contrast imaging solutions for space-based projects in visible/near-infrared at $10^{8}-10^{9}$ contrast levels. We developed several international collaborations to enable the characterization of coronagraphic components, wavefront sensor techniques, and correction solutions. We compared the chromatic behavior of several coronagraphs and found two coronagraphs that could be used for laboratory tests of chromatic corrections using the multi-DM  configuration of the bench: 2 deformable mirrors (DM), soon 3 DM. This capability also allows correcting both phase and amplitude aberrations, thus reaching the same contrast levels over the full-corrected field of view.

To reach very high contrast levels, environmental conditions of the laboratory bench must be optimized (optical quality, stabilization in temperature, etc) to simulate the stable space-based environment. Degrading these conditions by introducing an optical element creating an AO halo is easy. We plan to introduce residual atmospheric effects by implementing phase masks that simulate the AO residuals into a rotating wheel. These masks must be good enough in terms of quality to maintain a stable contrast level over the whole mask surface. These phase masks will allow us to study the impact of the averaging of the turbulent phase on the focal plane wavefront sensor but also the limitation on the data processing performance. Several AO contrast levels are considered to estimate the effect of the aberration level on the performance in the final images for both ELTs and 8m telescopes. We will also make sure that detection limit is still reached when the AO halo is removed. This is critical to ensure that no other limitations arise from the bench or the environment that would bias the laboratory study. Since the THD2 bench routinely reaches $10^{8}$ raw contrast and $10^{9}$ stability over a few minutes, it is well optimized for this study.

% References
\bibliography{Publi} % bibliography data in report.bib

\begin{thebibliography}{10}

\bibitem{Macintosh08}
{Macintosh}, B.~A., {Graham}, J.~R., {Palmer}, D.~W., {Doyon}, R., {Dunn}, J.,
  {Gavel}, D.~T., {Larkin}, J., {Oppenheimer}, B., {Saddlemyer}, L.,
  {Sivaramakrishnan}, A., {Wallace}, J.~K., {Bauman}, B., {Erickson}, D.~A.,
  {Marois}, C., {Poyneer}, L.~A., and {Soummer}, R., ``{The Gemini Planet
  Imager: from science to design to construction},'' in [{\em Society of
  Photo-Optical Instrumentation Engineers (SPIE) Conference
  Series}{\nolinebreak\hspace{0.1em}]},  {\em Society of Photo-Optical
  Instrumentation Engineers (SPIE) Conference Series} {\bf 7015} (July 2008).

\bibitem{Beuzit08}
{Beuzit}, J.-L., {Feldt}, M., {Dohlen}, K., {Mouillet}, D., {Puget}, P.,
  {Wildi}, F., {Abe}, L., {Antichi}, J., {Baruffolo}, A., {Baudoz}, P.,
  {Boccaletti}, A., {Carbillet}, M., {Charton}, J., {Claudi}, R., {Downing},
  M., {Fabron}, C., {Feautrier}, P., {Fedrigo}, E., {Fusco}, T., {Gach}, J.-L.,
  {Gratton}, R., {Henning}, T., {Hubin}, N., {Joos}, F., {Kasper}, M.,
  {Langlois}, M., {Lenzen}, R., {Moutou}, C., {Pavlov}, A., {Petit}, C.,
  {Pragt}, J., {Rabou}, P., {Rigal}, F., {Roelfsema}, R., {Rousset}, G.,
  {Saisse}, M., {Schmid}, H.-M., {Stadler}, E., {Thalmann}, C., {Turatto}, M.,
  {Udry}, S., {Vakili}, F., and {Waters}, R., ``{SPHERE: a planet finder
  instrument for the VLT},'' in [{\em Society of Photo-Optical Instrumentation
  Engineers (SPIE) Conference Series}{\nolinebreak\hspace{0.1em}]},  {\em
  Society of Photo-Optical Instrumentation Engineers (SPIE) Conference Series}
  {\bf 7014} (Aug. 2008).

\bibitem{Mazoyer14}
{Mazoyer}, J., {Baudoz}, P., {Galicher}, R., and {Rousset}, G.,
  ``{High-contrast imaging in polychromatic light with the self-coherent
  camera},'' {\em \aap}~{\bf 564},  L1 (Apr. 2014).

\bibitem{Delorme16_DZPM}
{Delorme}, J.~R., {N'Diaye}, M., {Galicher}, R., {Dohlen}, K., {Baudoz}, P.,
  {Caillat}, A., {Rousset}, G., {Soummer}, R., and {Dupuis}, O., ``{Laboratory
  validation of the dual-zone phase mask coronagraph in broadband light at the
  high-contrast imaging THD testbed},'' {\em \aap}~{\bf 592},  A119 (Aug.
  2016).

\bibitem{Shaklan06}
{Shaklan}, S.~B. and {Green}, J.~J., ``{Reflectivity and optical surface height
  requirements in a broadband coronagraph. 1.Contrast floor due to controllable
  spatial frequencies},'' {\em \ao}~{\bf 45},  5143--5153 (July 2006).

\bibitem{Singh14}
{Singh}, G., {Martinache}, F., {Baudoz}, P., {Guyon}, O., {Matsuo}, T.,
  {Jovanovic}, N., and {Clergeon}, C., ``{Lyot-based Low Order Wavefront Sensor
  for Phase-mask Coronagraphs: Principle, Simulations and Laboratory
  Experiments},'' {\em \pasp}~{\bf 126},  586 (June 2014).

\bibitem{Borde06}
{Bord{\'e}}, P.~J. and {Traub}, W.~A., ``{High-Contrast Imaging from Space:
  Speckle Nulling in a Low-Aberration Regime},'' {\em \apj}~{\bf 638},
  488--498 (Feb. 2006).

\bibitem{Baudoz06}
Baudoz, P., Boccaletti, A., Baudrand, J., and Rouan, D., ``The self-coherent
  camera: a new tool for planet detection,'' in [{\em IAU Colloq. 200: Direct
  Imaging of Exoplanets: Science Techniques}{\nolinebreak\hspace{0.1em}]},
  {Aime}, C. and {Vakili}, F., eds.,  553--558 (2006).

\bibitem{Galicher08}
{Galicher}, R., {Baudoz}, P., and {Rousset}, G., ``{Wavefront error correction
  and Earth-like planet detection by a self-coherent camera in space},'' {\em
  \aap}~{\bf 488},  L9--L12 (Sept. 2008).

\bibitem{Delorme16_MRSCC}
{Delorme}, J.~R., {Galicher}, R., {Baudoz}, P., {Rousset}, G., {Mazoyer}, J.,
  and {Dupuis}, O., ``{Focal plane wavefront sensor achromatization: The
  multireference self-coherent camera},'' {\em \aap}~{\bf 588},  A136 (Apr.
  2016).

\bibitem{Baudoz12}
{Baudoz}, P., {Mazoyer}, J., {Mas}, M., {Galicher}, R., and {Rousset}, G.,
  ``{Dark hole and planet detection: laboratory results using the self-coherent
  camera},'' in [{\em Ground-based and Airborne Instrumentation for Astronomy
  IV}{\nolinebreak\hspace{0.1em}]},  {\em \procspie} {\bf 8446},  84468C (Sept.
  2012).

\bibitem{Mazoyer13}
{Mazoyer}, J., {Baudoz}, P., {Galicher}, R., {Mas}, M., and {Rousset}, G.,
  ``{Estimation and correction of wavefront aberrations using the self-coherent
  camera: laboratory results},'' {\em \aap}~{\bf 557},  A9 (Sept. 2013).

\bibitem{Sauvage12}
{Sauvage}, J.-F., {Mugnier}, L., {Paul}, B., and {Villecroze}, R.,
  ``{Coronagraphic phase diversity: a simple focal plane sensor for
  high-contrast imaging},'' {\em Optics Letters}~{\bf 37},  4808 (Dec. 2012).

\bibitem{Herscovici17}
et~al., H.-S., ``Focal plane complex wavefront sensing with coronagraphic phase
  diversity on the thd2 bench,'' in [{\em AO for ELT 5
  Proceedings}{\nolinebreak\hspace{0.1em}]},  (2017).

\bibitem{Rouan00}
{Rouan}, D., {Riaud}, P., {Boccaletti}, A., {Cl{\'e}net}, Y., and {Labeyrie},
  A., ``{The Four-Quadrant Phase-Mask Coronagraph. I. Principle},'' {\em
  \pasp}~{\bf 112},  1479--1486 (Nov. 2000).

\bibitem{Soummer03}
{Soummer}, R., {Dohlen}, K., and {Aime}, C., ``{Achromatic dual-zone phase mask
  stellar coronagraph},'' {\em \aap}~{\bf 403},  369--381 (May 2003).

\bibitem{Hou14}
{Hou}, F., {Cao}, Q., {Zhu}, M., and {Ma}, O., ``{Wide-band six-region phase
  mask coronagraph},'' {\em Optics Express}~{\bf 22},  1884 (Jan. 2014).

\bibitem{Mawet09}
{Mawet}, D., {Serabyn}, E., {Liewer}, K., {Hanot}, C., {McEldowney}, S.,
  {Shemo}, D., and {O'Brien}, N., ``{Optical Vectorial Vortex Coronagraphs
  using Liquid Crystal Polymers: theory, manufacturing and laboratory
  demonstration},'' {\em Optics Express}~{\bf 17},  1902--1918 (Feb. 2009).

\end{thebibliography}
\bibliographystyle{spiebib} % makes bibtex use spiebib.bst

\end{document}